\documentclass[preprint]{aastex}

\usepackage{xspace}
\usepackage{multirow}

\newcommand{\Cstat}{$\mathcal{C}$-statistic\xspace}
\newcommand{\Fstat}{$\mathcal{F}$-statistic\xspace}
\newcommand{\iid}{\stackrel{\mathrm{i.i.d.}}{\sim}}

\slugcomment{Accepted for publication in The Astrophysical Journal}

\begin{document}

\title{Reanalysis of $\mathcal{F}$-statistic gravitational-wave searches with the higher criticism statistic}

\author{M. F. Bennett, A. Melatos}
\affil{School of Physics, University of Melbourne, Parkville, VIC 3010, Australia}
\email{mfb@unimelb.edu.au}
\and
\author{A. Delaigle, P. Hall}
\affil{Department of Mathematics and Statistics, University of Melbourne, Parkville, VIC 3010, Australia}

\begin{abstract}
	We propose a new method of gravitational wave detection using a modified form of higher criticism, a statistical technique introduced by \citet{don04}.  Higher criticism is designed to detect a group of sparse, weak sources, none of which are strong enough to be reliably estimated or detected individually.  We apply higher criticism as a second-pass method to synthetic \Fstat and \Cstat data for a monochromatic periodic source in a binary system and quantify the improvement relative to the first-pass methods.  We find that higher criticism on \Cstat data is more sensitive by $\sim 6$\% than the \Cstat alone under optimal conditions (i.e. binary orbit known exactly) and the relative advantage increases as the error in the orbital parameters increases.  Higher criticism is robust even when the source is not monochromatic (e.g. phase wandering in an accreting system).  Applying higher criticism to a phase-wandering source over multiple time intervals gives a $\gtrsim 30$\% increase in detectability with few assumptions about the frequency evolution.  By contrast, in all-sky searches for unknown periodic sources, which are dominated by the brightest source, second-pass higher criticism does not provide any benefits over a first pass search. \end{abstract}


\section{Introduction}\label{sec:intro}
Direct detection of gravitational waves appears likely in the near future.  Existing terrestrial long-baseline interferometers, such as the Laser Interferometer Gravitational-Wave Observatory (LIGO) and Virgo, have achieved their design sensitivity \citep{abb09c, acc12}.  Next-generation interferometers now under construction are expected to detect tens of events per year, if contemporary estimates of compact binary coalescence rates are correct \citep{aba10a}.

Searches for periodic sources have the advantage of integrating over long observation times to increase the signal-to-noise ratio.  The most likely periodic sources detectable by terrestrial interferometers are rapidly rotating, slightly deformed neutron stars \citep{bil98, ush00, mel05}.  They emit at the spin frequency $f_*$ and its first harmonic $2f_*$ \citep{jar98}.  X-ray timing measurements find 0.3 kHz $\lesssim f_* \lesssim$ 0.6 kHz for low-mass X-ray binaries \citep{cha03}, placing these sources squarely in the LIGO-Virgo band.  Directed searches for electromagnetically observed targets \citep{abb04, abb07a, abb08b, aba10b, aba11} and blind, all-sky searches for unknown sources \citep{abb05, abb07b, abb08a, abb09a, abb09b, abb09d} have both been reported.

Targeted searches for known pulsars use the radio ephemeris to guide the search and reduce computational expense by assuming the electromagnetic and gravitational wave phases track each other closely.  One search for the Crab pulsar allowed for a mismatch of up to one part in $10^{4}$ between twice the radio pulse frequency and the gravitational wave frequency \citep{abb08b}.  Targeted searches for the Crab and Vela pulsars have set direct upper limits on the wave strain of $3.4 \times 10^{-25}$ and $2.2 \times 10^{-24}$ respectively, beating the indirect spin-down limits inferred from radio observations \citep{abb08b, aba11}.  A search of 78 pulsars using data from the third and fourth science runs of the LIGO and GEO 600 detectors set upper limits on the wave strain $h$ and ellipticity $\varepsilon$, the tightest of which are $h < 2.6 \times 10^{-25}$ for PSR J1603-7202 and $\varepsilon < 7.1 \times 10^{-7}$ for PSR J2124-3358 \citep{abb07a}.

Blind, all-sky searches address more sources than targeted searches but are expensive computationally as they cover a larger parameter domain to keep track of the unknown frequency evolution.  A number of LIGO all-sky searches for periodic sources have been conducted \citep{abb05, abb07b, abb08a, abb09a, abb09b, abb09d}, some of which leverage the distributed computing power of the Einstein@Home project \citep{abb09a, abb09d}.  Frequently, they are based on a maximum-likelihood detection statistic, called the \Fstat \citep{jar98}.  The \Fstat is computed by coherently integrating over the observation time $T_{obs}$, assuming a biaxial neutron star at a specific spin frequency and sky position.  \citet{cut05} generalized the \Fstat to apply to multiple detectors and sources.

The computational expense of a fully coherent search for unknown sources becomes prohibitive as $T_{obs}$ increases.  Semi-coherent methods address this problem, dividing the observation time into intervals, which are individually searched coherently but combined incoherently.  Semi-coherent methods are more sensitive than a fully coherent search for the same computational cost if the parameter space is large \citep{wet12}.  \citet{abb08a} reported results of a semi-coherent all-sky search for periodic sources.  They described and compared three semi-coherent methods:  StackSlide, PowerFlux, and Hough number count \citep{bra00, kri04, cut05b, abb08a}.  For binary sources, \citet{mes07} derived the \Cstat, which incoherently combines individual \Fstat values with a comb template that matches the orbital sideband structure.

In this paper, we propose a new method for enhancing gravitational wave searches using a second-pass method known as higher criticism.  Higher criticism was suggested originally by John Tukey and developed further by \citet{don04}, who proved that it has some mathematical optimality properties in the case of a specific noise distribution.    The ultimate goal of any gravitational wave search is to identify an individual object as a definite source.  Higher criticism is not suited for this purpose.  It is designed to search against a known background for a group of sparse signals too weak to be detected individually.  Higher criticism detects the presence of the group but cannot reliably estimate the waveform.  We discuss applications of higher criticism to gravitational wave detection, focusing on the \Fstat \citep{jar98} as a case study.  In general, we find that higher criticism is significantly more robust than other methods; it ignores some information that other detection statistics incorporate, so may be less sensitive, but makes fewer assumptions about the form of the signal, especially important where the signal evolves during the search.

The paper is structured as follows.  In Section \ref{sec:HC}, we define the higher criticism statistic, compare the detectability of signals with different amplitudes and sparsities, and compute detection thresholds.  In Section \ref{sec:Fstat}, we briefly review the \Fstat.  In Section \ref{sec:binary}, we apply a form of higher criticism to targeted searches for a binary pulsar and compare its performance with the \Cstat.  We discuss how higher criticism can accommodate phase wandering in Section \ref{sec:wandering}.  While we apply higher criticism specifically to periodic gravitational wave searches in this paper, it is a general method with other applications, including detecting non-Gaussianity in Wilkinson Microwave Anisotropy  Probe (WMAP) data \citep{cay05}.

\section{Higher Criticism}\label{sec:HC}

\subsection{Theory}\label{sec:HCtheory}
The higher criticism statistic was introduced by \citet{don04} to test the hypothesis that $n$ independent and identically distributed (i.i.d.) samples $X_1, \ldots,X_n$ come from the same zero-mean Gaussian distribution $N(0, 1)$, against the alternative that a small fraction of them have a nonzero mean $\mu>0$, that is, to test:
\begin{eqnarray}
	H_0 &:& X_i \iid N(0,1)~\ \ \ \  \textrm{ for \ $i=1,\ldots,n$}~,\\
	H_1 &:& X_i \iid (1 - \epsilon) N(0,1) + \epsilon N(\mu, 1)~\ \ \ \  \textrm{ for \ $i=1,\ldots,n$}~.\label{eq:H1}
\end{eqnarray}
For example,  $H_0$ could correspond to a situation where all samples are just background noise with no signal ($\mu=0$), whereas in $H_1$ a fraction $\epsilon$ of samples  contain signal ($\mu\neq 0$).  Higher criticism is particularly designed for the very challenging situation where the signal is sparse ($\epsilon\ll 1$) and weak ($\mu$ is so small that the largest extreme values under $H_1$ are essentially the same as those under $H_0$).  Although we introduce the statistic in the context of Gaussian noise, it can be generalized easily to other distributions (see Section \ref{sec:generalHC}). The case of correlated, or non-white, noise has been treated by \citet{hal10}.

Let $Z$ denote a random variable with distribution $N(0,1)$. The higher criticism statistic $HC$ is computed from the $n$ p-values $p_i = P[Z > X_i]$ generated by the $n$ tests $H_{0,i}: X_i \iid N(0,1)$ against $H_{1,i}: X_i \iid N(\mu,1)$, $\mu>0$, $i=1,\ldots,n$. It is defined by \citep{don04},
\begin{equation}\label{eq:HCdef}
	HC = \max_{1 \le i \le n} \frac{\sqrt{n} [i/n - p_{(i)}]}{\sqrt{p_{(i)}[1 - p_{(i)}]}}~,
\end{equation}
where $p_{(1)}\leq p_{(2)}\leq \cdots\leq p_{(n)}$ are the $p_i$'s sorted into increasing order.
The statistic $HC$ rejects $H_0$ when $HC>g(n, \alpha)$, where $g(n, \alpha)$ is a threshold chosen so that $P_{H_0}[HC>g(n, \alpha)]=P_{H_0}(\textrm{reject $H_0$})\leq \alpha$, where $P_{H_0}$ denotes the probability when $H_0$ is true.
[Our notation differs from \citet{don04} in that we use the symbol $g$ for the threshold instead of $h$ to avoid confusion with the gravitational wave strain $h$.]
In other words, the higher criticism test detects a signal when $HC > g(n, \alpha)$.
\citet{don04} showed that, when $H_0$ is true, one has $HC \sim \sqrt{2 \log\log(n)}$ asymptotically as $n\to\infty$ for all $\alpha > 0$, and that if $0<\alpha<1$ and $n$ is large enough, one can use the threshold $g(n, \alpha) \approx \sqrt{2 \log\log(n)}$. However for more general $\alpha$ and $n$, this does not work well in practice, as we illustrate in Section \ref{sec:detectthresh}, where we discuss how to choose $g(n, \alpha)$ accurately as a function of $n$ and $\alpha$ for finite $n$.

\begin{figure}
	\plotone{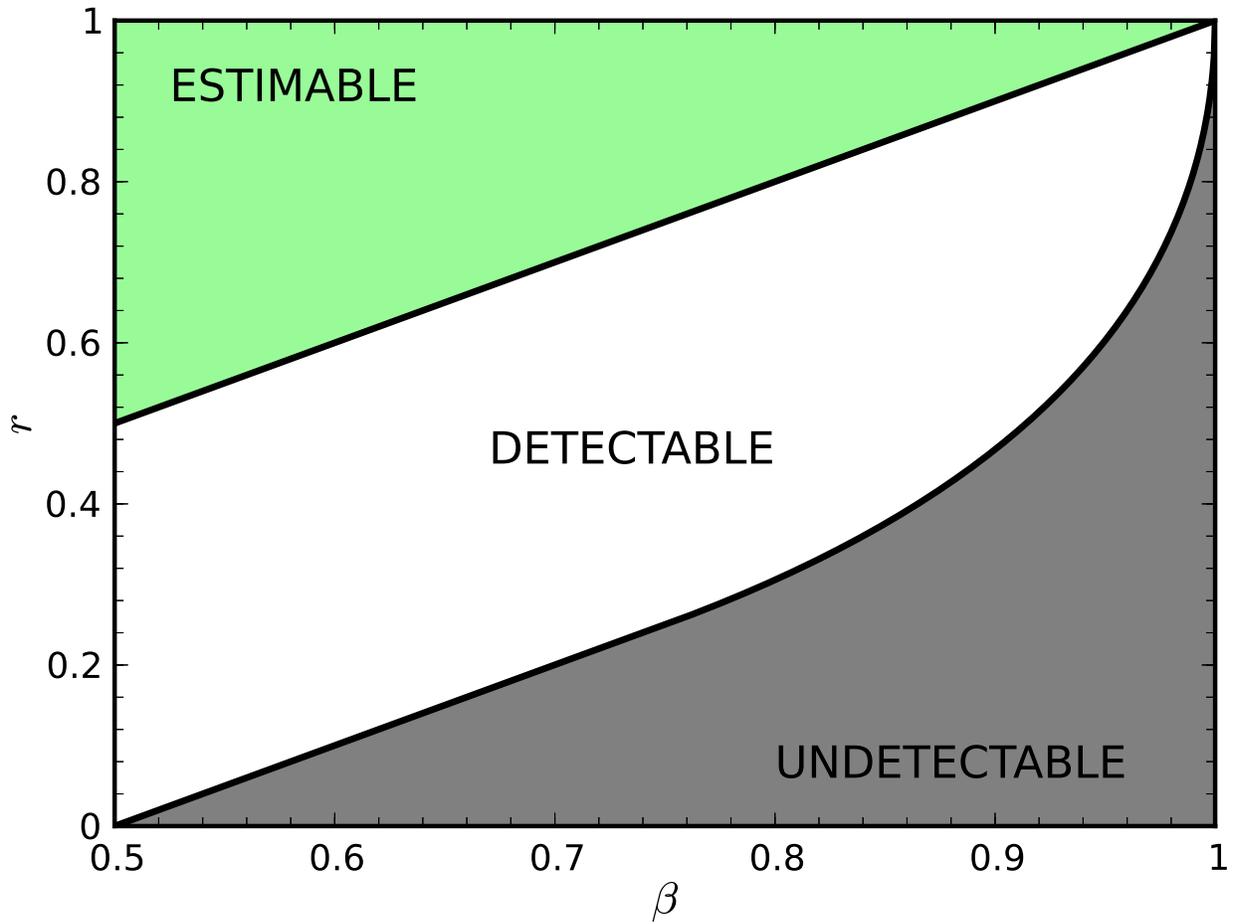}
	\caption{Detectability regions on the $r$-$\beta$ plane.  Undetectable region, shaded gray; detectable region, where signal can be detected but not estimated, unshaded; estimable region, where signal can be both detected and estimated, shaded green.\label{fig:detectregions}}
\end{figure}

Next, we quantify more precisely how small $\epsilon$ and $\mu$ can be for the test based on $HC$ to be able to distinguish $H_0$ from $H_1$.
Let $r$ and $\beta$ be two positive parameters. \citet{don04} study the properties of the $HC$ statistic for $\mu$ and $\epsilon$ of the form
\begin{eqnarray}
	\mu &=& \sqrt{2 r \log n}~,\label{eq:mu}\\
	\epsilon &=& n^{-\beta}\label{eq:parameps}~.
\end{eqnarray}
These represent very difficult situations since, with $r$ small, $\mu$ of this form is smaller than the mean of the upper extreme statistics, and with $\beta$ large, the proportion of samples with non zero mean is very close to zero.
Clearly, a detection is not possible if $r$ is very small and $\beta$ is very large at the same time. \citet{don04} showed that, as $n\to\infty$, there exists a function $\rho(\beta)$ defined by
\begin{equation}\label{eq:detection_boundary}
	\rho(\beta) = \Bigg\{ \begin{array}{ll} \beta - 1/2~, & \quad 1/2 < \beta \le 3/4~, \\ (1 - \sqrt{1 - \beta})^2~, & \quad 3/4 < \beta < 1~,\end{array}
\end{equation}
such that $H_0$ and $H_1$ are distinguishable only if $r > \rho(\beta)$. If $r < \rho(\beta)$, there does not exist a test that can distinguish $H_0$ from $H_1$.
\citet{don04} proved that, under some conditions, the $HC$ statistic is able to distinguish $H_0$ from $H_1$  throughout the detectable region $r > \rho(\beta)$, and that it has full asymptotic power, that is, $P_{H_1}(\textrm{reject $H_0$}) \to 1$ as $n\to\infty$ (here $P_{H_1}$ denotes the conditional probability when $H_1$ is true).
Another test has also been considered in the literature, namely the Neyman-Pearson likelihood ratio test, but it is less attractive since it requires  $r$ and $\beta$ to be known \citep{don04}, whereas the test based on the higher criticism statistic does not.
Interestingly, \citet{don04} also show that, in addition to the detection boundary, there is a second region $r>\beta$ called the estimable region, where $H_0$ and $H_1$ can be distinguished and the mean $\mu$ can also be estimated consistently.  Figure \ref{fig:detectregions}, which can be found in \citet{don04}, illustrates the detectable and estimable regions on the $r$-$\beta$ plane.

\subsection{Distribution of $HC$}
In this section, we present briefly some results from Monte-Carlo simulations to illustrate the behavior of $HC$ as a function of signal amplitude $\mu$ and sparsity $\epsilon$ for $n=10^6$ sample values drawn randomly from the $H_1$ distribution given by equation (\ref{eq:H1}).

\begin{figure}
	\plotone{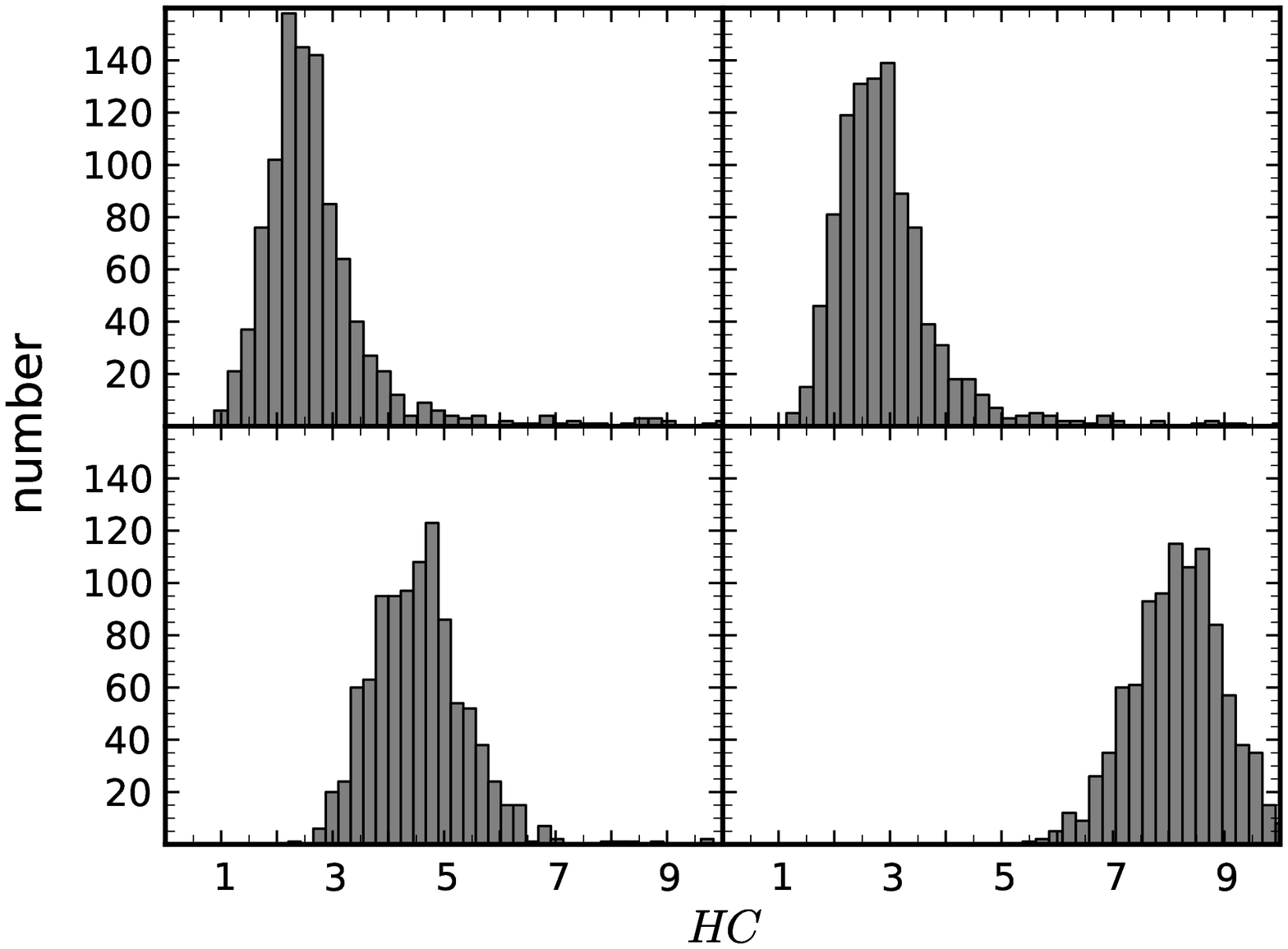}
	\caption{Histograms of $HC$ for $10^3$ Monte-Carlo trials, each constructed from $n=10^6$ samples drawn from distribution $H_1$ with sparsity $\epsilon=5 \times 10^{-3}$ and increasing amplitude $\mu$.  Upper left panel: noise only ($\mu = 0$), i.e. $H_0$ distribution.  Remaining panels:  $\mu = 0.1, 0.3, 1$ (upper right, lower left, lower right). \label{fig:HCdist-gaussian-vary_mean}}
\end{figure}

\begin{figure}
	\plotone{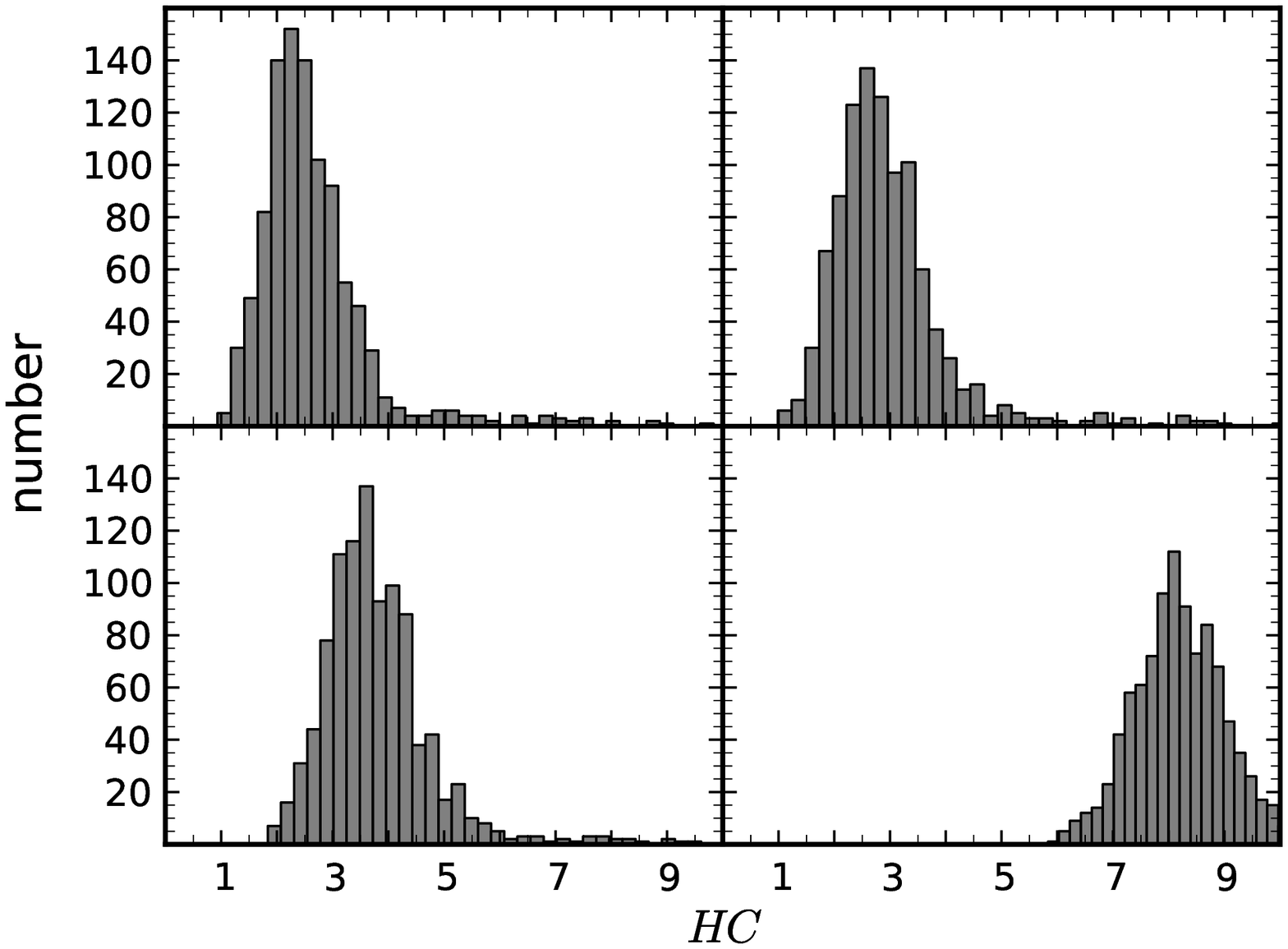}
	\caption{Histograms of $HC$ for $10^3$ Monte-Carlo trials, each constructed from $n=10^6$ samples drawn from distribution $H_1$ with amplitude $\mu=1$ and increasing sparsity $\epsilon$.  Upper left panel: noise only ($\epsilon = 0$), i.e. $H_0$ distribution.  Remaining panels:  $\epsilon = 5 \times 10^{-4}, 2 \times 10^{-3}, 5 \times 10^{-3}$ (upper right, lower left, lower right). \label{fig:HCdist-gaussian-vary_sparsity}}
\end{figure}

Figure \ref{fig:HCdist-gaussian-vary_mean} displays a histogram of $HC$ values for $10^3$ trials with $\epsilon = 5 \times 10^{-3}$ and $\mu = 0, 0.1, 0.3, 1$.  Figure \ref{fig:HCdist-gaussian-vary_sparsity} displays a similar histogram with $\mu=1$ and $\epsilon = 0, 5 \times 10^{-4}, 2 \times 10^{-3}, 5 \times 10^{-3}$.  The top left panel in both figures consists of samples drawn from the null distribution $H_0$.  As $\mu$ or $\epsilon$ increases, so too does the average value of $HC$.

\subsection{$HC$ for a general noise distribution}\label{sec:generalHC}
In Section \ref{sec:HCtheory}, we defined $HC$ in the context of Gaussian noise.  As shown in \citet{don04}, higher criticism can be generalized  to other noise and signal distributions, $\mathcal{N}$ and  $\mathcal{S}(\mu)$ say, where $\mu$ describes the amplitude of the signal (e.g. mean of the Gaussian distribution in Section \ref{sec:HCtheory}). In this more general setting we test to discriminate between the hypotheses
\begin{eqnarray}
	H_0 &:& X_i \iid \mathcal{N} \ \ \ \  \textrm{ for \ $i=1,\ldots,n$}~, \\
	H_1 &:& X_i \iid (1 - \epsilon) \mathcal{N} + \epsilon \mathcal{S}(\mu)\ \ \ \  \textrm{ for \ $i=1,\ldots,n$}~.
\end{eqnarray}
In this context, the higher criticism statistic is computed as in equation (\ref{eq:HCdef}), but replacing the $p_{(i)}$'s by the ordered values of $p_i = P[Z > X_i]$, $i=1,\ldots,n$, where $Z$ is a random variable that has the distribution $\mathcal{N}$; see also \citet{DH09}.

The detection boundary varies with the noise distribution. For example, when  $\mathcal{N}$ is the $\chi^2$ distribution that we use later, the proportion $\epsilon$ is defined as in equation (\ref{eq:parameps}), but the intensity of the signal is defined in terms of the noncentrality parameter $\rho^2$ of the $\chi^2$ distribution, through
\begin{equation}\label{eq:paramchi2}
	\rho^2 = 2 r \log n~
\end{equation}
instead of the mean $\mu$ in  equation (\ref{eq:mu}). With these definitions of $r$ and $\beta$, the detection boundary is given by equation
(\ref{eq:detection_boundary}), as in the normal case.

\subsection{Detection thresholds}\label{sec:detectthresh}
In this section, we discuss the choice of the detection threshold $g(n, \alpha)$ for $HC$. As already noted earlier, the threshold satisfies $P_{H_0}[HC > g(n, \alpha)]=P_{H_0}(\textrm{reject }H_0) \le \alpha$, and under somewhat restrictive conditions, it can be approximated asymptotically by $\sqrt{2 \log\log(n)}$. However, in finite samples and under more general conditions, this theoretical value $\sqrt{2 \log\log(n)}$ is not a good approximation to $g(n, \alpha)$.

To illustrate this we conduct Monte Carlo simulations to determine the threshold $g(n, \alpha)$ for finite $n$. Note that, when $H_0$ is true, the p-values are always independent and identically distributed according to the uniform distribution $U(0,1)$, and thus $g(n, \alpha)$ is independent of the specific noise distribution $\mathcal{N}$. Therefore it suffices to generate p-values from the $U(0,1)$ distribution to compute the threshold.
We run $10^6$ noise-only simulations with $n = 10^3, 10^4, 10^5, 10^6$ to determine Monte-Carlo thresholds.  For each, we generate $n$ p-values from $U(0,1)$ and compute $HC$.

\begin{figure}
	\plotone{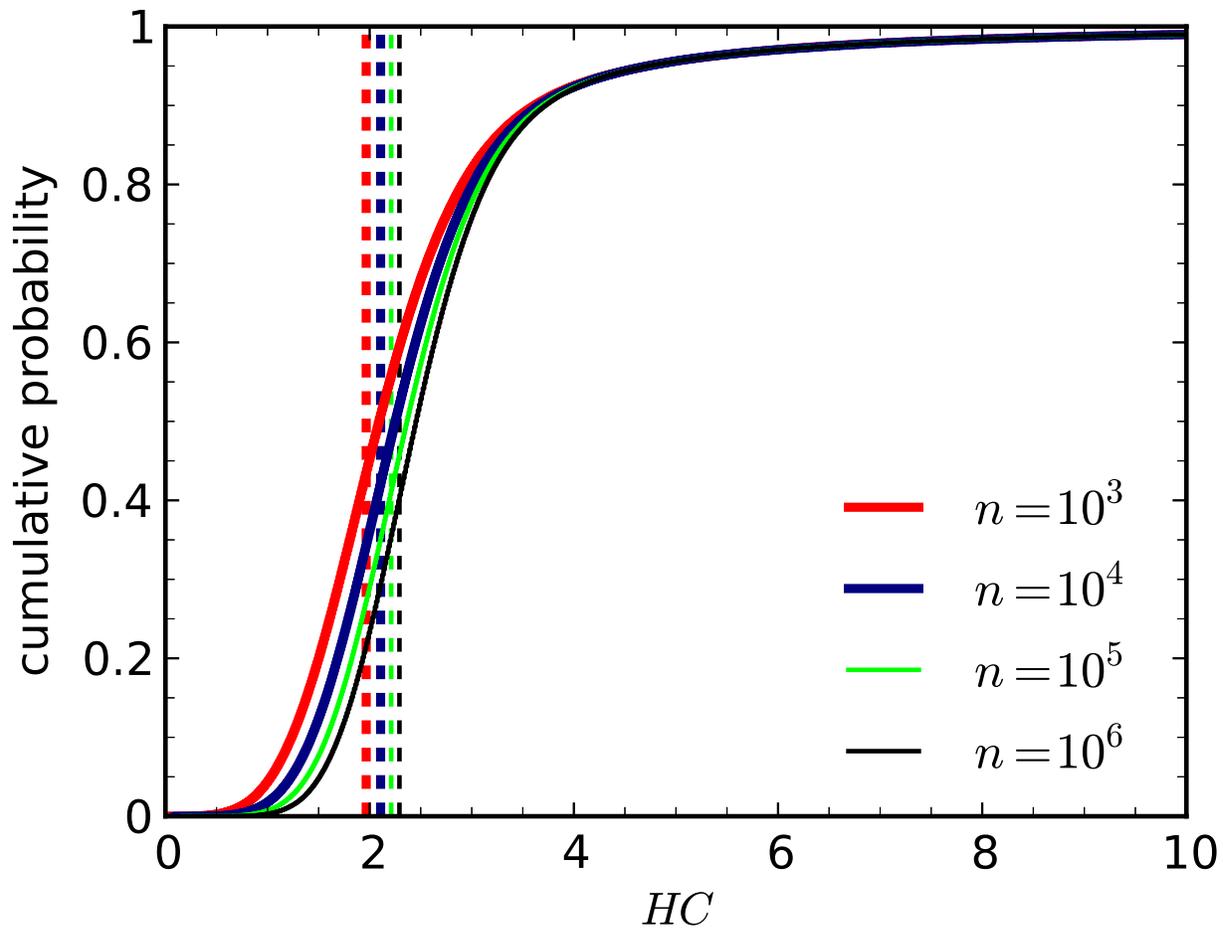}
	\caption{Cumulative distribution function of $HC$ when $n=10^3$ (red), $n=10^4$ (blue), $n=10^5$ (green), and $n=10^6$ (black) samples drawn from $H_0$.  Dashed vertical lines indicate the corresponding theoretical $n\to\infty$ threshold $g(n, \alpha) = \sqrt{2 \log\log(n)}$.\label{fig:CDF}}
\end{figure}

Figure \ref{fig:CDF} displays the cumulative distribution function of $HC$ for the simulated noise samples.
Threshold values obtained from the simulations by solving an empirical version of $P_{H_0}(HC > g(n, \alpha)) \approx \alpha$ are listed in Table \ref{tab:thresholds} for $n=10^3, 10^4, 10^5, 10^6$ and $\alpha=0.5, 0.1, 0.05, 0.01$.
For comparison, the  theoretical value $\sqrt{2 \log\log(n)}$ is also quoted.  From this and Figure \ref{fig:CDF}, we can see that this theoretical value generally does not work well in practice.  In each case, a large fraction ($\gtrsim$ 50\%) of noise-only trials fall above this threshold.  In the remainder of this paper, we commonly use the threshold $g(n, 0.01)$ as determined from Monte-Carlo simulations with $\alpha=0.01$ (1\% false alarm rate).
\begin{table}
\begin{center}
\caption{$HC$ threshold $g(n, \alpha)$. \label{tab:thresholds}}
\begin{tabular}{ccccc}
	\tableline\tableline
	False alarm rate & \multicolumn{4}{c}{$HC$ threshold $g(n, \alpha)$} \\
	$\alpha$ & $n=10^3$ & $n=10^4$ & $n=10^5$ & $n=10^6$ \\
	\tableline
	0.5  & 2.10 & 2.26 & 2.37 & 2.46 \\
	0.1  & 3.62 & 3.66 & 3.70 & 3.73 \\
	0.05 & 4.72 & 4.73 & 4.73 & 4.73 \\
	0.01 & 10.0 & 10.1 & 10.1 & 10.0 \\
	\tableline
	Theoretical threshold & \multirow{2}{*}{1.97} & \multirow{2}{*}{2.11} & \multirow{2}{*}{2.21} & \multirow{2}{*}{2.29} \\
	$\sqrt{2\log\log(n)}$ & & & \\
	\tableline
\end{tabular}
\end{center}
\end{table}

\section{$\mathcal{F}$-statistic}\label{sec:Fstat}
The \Fstat is an efficient detection statistic for periodic gravitational waves based on maximum likelihood.  A number of blind and targeted searches for periodic sources have been carried out using it and its relatives \citep{abb04, abb05, abb07b, abb08a, abb08b, abb09a, abb09d, aba10b, aba11}.  We use the notation $\chi^2_k(\lambda)$ to represent the noncentral (central if $\lambda=0$) $\chi^2$ distribution with $k$ degrees of freedom and noncentrality parameter $\lambda$.  In the absence of a signal, and assuming stationary, Gaussian noise, $2 \mathcal{F}$ is distributed according to a $\chi^2_4(0)$ distribution, while in the presence of a signal, $2 \mathcal{F}$ obeys a $\chi^2_4(\rho^2)$ distribution \citep{jar98, abb07b}.  In terms of the waveform, one writes
\begin{equation}\label{eq:noncentrality_paramater}
	\rho^2 = \frac{2}{S_h(f)} \int_0^{T_{obs}} dt \, h(t)^2~,
\end{equation}
where $S_h(f)$ is the spectral noise density of the interferometer, $T_{obs}$ is the total observation time, and $h(t)=F_+(t)h_+(t)+F_\times(t)h_\times(t)$ is the gravitational wave strain at the detector, which depends on the beam pattern functions $F_+$ and $F_\times$.  For a biaxial star, the strains $h_+(t)$ and $h_\times(t)$ depend on the wobble angle $\theta$ between the rotation and symmetry axes, the inclination angle $\iota$ between the rotation axis and line of sight, right ascension $\alpha_{\mathrm{sky}}$, declination $\delta_{\mathrm{sky}}$, polarization angle $\psi$, and the characteristic strain amplitude
\begin{equation}
	h_0 = \frac{16 \pi^2 G}{c^4} \frac{\varepsilon I f^2}{D}~,
\end{equation}
where $\varepsilon$ is the ellipticity, $I$ is the moment of inertia, and $D$ is the distance to the star \citep{jar98}.  We obtain an approximate relation between $\rho^2$ and $h_0$ by averaging equation (\ref{eq:noncentrality_paramater}) over all relevant angles \citep{jar98, vig09},
\begin{equation}\label{eq:rho2h0relation}
	\rho^2 \approx \frac{32}{375} \frac{h_0^2 T_{obs}}{S_h}~,
\end{equation}
with the assumption that the spectral noise density is achromatic, viz. $S_h(f) = S_h$.  Equation (\ref{eq:rho2h0relation}) allows us to translate detection limits on $\rho^2$ into limits on $h_0$ throughout the rest of the paper.

For a known source, the amplitude of the average signal that can be detected coherently with a 1\% false alarm rate and a 10\% false dismissal rate is \citep{wet12}
\begin{equation}\label{eq:LIGOthresh_averaged}
	h_0 = 15.6 \sqrt{\frac{S_h(f)}{T_{obs}}}~.
\end{equation}
Equation (\ref{eq:LIGOthresh_averaged}) is often seen with a factor of 11.4 instead of 15.6 \citep[e.g.,][]{abb07b}.  The 11.4 factor is obtained assuming the wobble angle $\theta = \pi/2$, whereas we have averaged over $\theta$ to arrive at equation (\ref{eq:LIGOthresh_averaged}).  For an unknown source, we search across a large number of bins and $h_0$ must be larger than the value in equation (\ref{eq:LIGOthresh_averaged}) to rise above the background.

\citet{cut05} derived the \Fstat for multiple detectors or multiple sources and found that combining the \Fstat values from multiple sources increases their detectability as long as the sources have a squared signal-to-noise ratio greater than one fifth of the brightest source.  For example, of the known millisecond pulsars, it is advantageous to combine the five strongest ones but no more.

Semi-coherent searches divide the observation time into $N$ intervals of length $\Delta T = T_{obs} / N$.  The \Fstat is computed for each time interval $i$ and combined to obtain
\begin{equation}
	2\mathcal{F}_{sc} = \sum_{i=1}^N 2\mathcal{F}_i~.
\end{equation}
As each $2\mathcal{F}_i$ follows a $\chi^2_4$ distribution, $2\mathcal{F}_{sc}$ follows a $\chi^2_{4N}$ distribution.

Searches for periodic sources require detailed knowledge of the frequency evolution.  For pulsars, the radio ephemeris is used to guide the search.  Any difference between the radio and gravitational wave phases causes significant problems.  For unknown sources, we have even less information and are powerless to correct for phase wandering.  Recently, \citet{cut11} proposed a ``phase-relaxed'' \Fstat for an all-sky search, modifying the semi-coherent detection statistic to accommodate a phase offset $\delta_i$ for each coherent time interval $i$, while preserving a monochromatic phase model overall.

In this paper we outline a new approach: applying higher criticism on a second pass to reanalyze the results of gravitational wave searches.  Higher criticism relies on detailed knowledge of the background noise distribution.  The detection statistics discussed so far assume stationary, Gaussian noise.  Realistic detector noise is neither stationary nor Gaussian.  We discuss this key point further in Section \ref{sec:discussion}.  For now, we persevere with the common, simplifying assumption of stationary, Gaussian detector noise in order to assess the overall viability of higher criticism, noting that modified forms of higher criticism have been constructed for correlated noise \citep{hal10} or when the noise distribution is imperfectly known \citep{del11}.

\section{Targeted binary search}\label{sec:binary}
Low-mass X-ray binaries (LMXBs) are accreting neutron stars in a binary orbit with a low mass companion.  X-ray pulsations and burst oscillations place the spin frequencies of discovered objects in the range 270 Hz $\le f_* \le$ 619 Hz \citep{cha03}, well below the centrifugal break-up frequency \citep{coo94}.  Gravitational radiation is one effective means to balance the accretion torque and stall spin up \citep{bil98}.  In this section, we compare a second-pass search with higher criticism to the side-band \Cstat algorithm proposed by \citet{mes07}.

\subsection{\Fstat and \Cstat}
The gravitational wave signal from a neutron star in a binary has its frequency modulated into sidebands with spacing $1/P$, where $P$ is the orbital period \citep{mes07}.  Therefore, an \Fstat search sees a large number of relatively weak signals across many frequency bands.  For a neutron star with intrinsic gravitational wave frequency $f_0$, $2\mathcal{F}$ obeys a $\chi^2_4[\rho^2(f)]$ distribution, where the frequency dependent noncentrality parameter can be written \citep{mes07}
\begin{equation}\label{eq:2Fapprox}
	\rho^2(f) = \rho_0^2 \sum_{n=-\lfloor{Z_0}\rfloor}^{\lfloor{Z_0}\rfloor} J_n^2(Z_0) \left|\tilde{W}(f-f_{0,n})\right|^2~,
\end{equation}
with $Z_0 = 2 \pi f_0 a$, where $\lfloor{Z_0}\rfloor$ represents the largest integer less than $Z_0$, $J_n$ is the $n$-th order Bessel function of the first kind, $a$ is the light crossing time of the orbital radius projected onto the line of sight, $f_{0,n} = f_0 - n/P$ is the frequency of the $n$th sideband around a source with frequency $f_0$, $\rho_0^2$ is the source signal-to-noise ratio squared [see equation (\ref{eq:noncentrality_paramater})], and $\tilde{W}(f)$ is the Fourier transform of the window function $W(t)$, which equals 1 or 0 when the detector is on or off respectively.  Assuming no gaps in the data, one finds
\begin{equation}\label{eq:windowfunction}
	\left| \tilde{W}(f) \right|^2 = \frac{\sin^2(\pi f T_{obs})}{\pi^2 f^2 T_{obs}^2}~.
\end{equation}
With the window function defined as in \citet{mes07}, an extra factor of $1/T_{obs}^2$ is included in equation (\ref{eq:windowfunction}) so that $\mathcal{C}$ is dimensionless.

The \Cstat sums incoherently the \Fstat power in the orbital sidebands \citep{mes07} as follows:
\begin{equation}\label{eq:C_f}
	\mathcal{C}(f) = \sum_{n=-\lfloor{Z_0}\rfloor}^{\lfloor{Z_0}\rfloor} 2 \mathcal{F}(f - n/P)~.
\end{equation}
For noise only, $2 \mathcal{F}$ is distributed according to a $\chi^2_4$ distribution.  Hence, $\mathcal{C}$, which is the sum of $M\;$ $2\mathcal{F}$ values, obeys a $\chi^2_{4M}$ distribution, where $M = 2\lfloor{Z_0}\rfloor + 1$ is the total number of sidebands \citep{mes07}.  In the presence of a signal, $2 \mathcal{F}(f)$ is distributed according to a noncentral $\chi^2_4[\rho^2(f)]$ distribution with a frequency dependent noncentrality parameter $\rho^2(f)$ and $\mathcal{C}$ follows a noncentral $\chi^2_{4M}(\lambda)$ distribution with noncentrality parameter \citep{mes07}
\begin{equation}\label{eq:lambda}
	\lambda = \sum_{n=-\lfloor{Z_0}\rfloor}^{\lfloor{Z_0}\rfloor} \rho^2(f - n/P)~.
\end{equation}

\subsection{Higher criticism}\label{sec:binaryHC}
How does a \Cstat search, which assumes that $P$ (and hence the sideband locations) are known, perform in comparison with a second-pass higher criticism search over \Fstat or \Cstat values?  We run Monte-Carlo simulations to answer this question.

We simulate a targeted search for Sco X-1, similar to the one described in 
Sammut et al. (2013, in preparation)
.  The search parameters are $P = 68023.84$ s, $a = 1.44$ s, $T_{obs} = 10$ days, and source frequency $f_0 = 400$ Hz (assumed).  We search over the range of frequencies $100 \le f \le 1000$ Hz with frequency bin spacing $\delta f = 1 / (2T_{obs}) = 5.8\times 10^{-7}$ Hz, corresponding to $n = 1.5 \times 10^9$ frequency bins in total.  The signal is modulated into $M$ sidebands separated by $1/P$.  $M$ depends on the signal frequency, with $1811 \le M \le 18097$ for 100 Hz $\le f_0 \le$ 1000 Hz.  The width in frequency space of the entire sideband structure (the comb) is 0.03 Hz $\le M/P \le$ 0.27 Hz.

Let us apply higher criticism to the problem.  We take advantage of the relatively narrow sideband structure and divide the frequency domain $[f_{min}, f_{max}]$ into $N_w$ windows of equal width $w$, each containing $n_w$ frequency bins, viz.  $[f_{min}, f_{min}+w]$, $[f_{min}+w/2, f_{min}+3w/2]$, $[f_{min}+w, f_{min}+2w]$, ..., $[f_{max}-3w/2, f_{max}-w/2]$, $[f_{max}-w, f_{max}]$.  The width $w$ is chosen to be twice the maximum width of the sidebands, $w = 2 M / P$, and windows overlap by 50\% to ensure that the entire signal is contained entirely within a single window.  A more sophisticated method would vary the width with frequency as the sideband width increases from 0.03 Hz for $f_0 = 100$ Hz to 0.27 Hz for $f_0 = 1000$ Hz, however, for simplicity, we construct our search windows based on the maximum sideband width at $f_{max} = 1000$ Hz, obtaining $w = 0.53$ Hz, $n_w = 9.2 \times 10^5$, and $N_w = 3.4 \times 10^3$.

The data are synthesized as follows.  First, the noncentrality parameter is computed for each frequency bin.  The signal is modulated into sidebands and the frequency dependent noncentrality can be written 
(Sammut et al. 2013, in preparation)
,
\begin{equation}
	\rho^2(f) \approx \rho_0^2 \sum_{n=-\lfloor{Z_0}\rfloor}^{\lfloor{Z_0}\rfloor} J_n^2(Z_0) \left|\tilde{W}(f-f_{0,n})\right|^2~.
\end{equation}
We simplify this equation assuming $\left|\tilde{W}(f)\right|^2 \approx \delta(f)$ (true in the limit $T_{obs} \to\infty$) and assume the $J_n^2(Z_0)$ all have similar amplitude, which we absorb into the constant $\rho_0^2$.  Additionally, we adjust the frequency of each sideband $f_{0,n}$ to be equal to the corresponding closest frequency bin, $\textrm{nr}(f_{0,n})$, using $\textrm{nr}(f)$ to represent the frequency bin closest to frequency $f$, neglecting the potential loss of amplitude caused by the mismatch between $f_{0,n}$ and $\textrm{nr}(f_{0,n})$.   We denote the frequency on the $i$-th frequency bin $f'_i$, using the `prime' notation to differentiate between frequency bins and the exact frequency of the signal $f_0$ or the sidebands $f_{0,n}$.
   The simplified noncentrality parameter used to generate the \Fstat for each frequency bin $f'_i$ is
\begin{equation}\label{eq:simplerho2}
	\rho^2(f'_i) = \rho_0^2 \sum_{n=-\lfloor{Z_0}\rfloor}^{\lfloor{Z_0}\rfloor} K\left[f'_i, \textrm{nr}(f_{0,n})\right]~,
\end{equation}
where $K(x, y) = \delta_{x,y}$ is the standard Kronecker delta rebranded so that we can see the subscripts clearly.  Equation (\ref{eq:simplerho2}) corresponds to $\rho^2(f'_i)$ equal to $\rho_0^2$ for the frequency bin closest to each of the true sideband locations and zero elsewhere.  We generate synthetic $2\mathcal{F}$ values drawn from the $\chi^2_4[\rho^2(f'_i)]$ distribution, for each frequency bin $f'_i$.

To compute \Cstat values, we construct a comb centered at each frequency bin $f'_i$ assuming $f_0 = f'_i$,
\begin{equation}
	q(f'_i) = \sum_{n=-\lfloor{Z'_i}\rfloor}^{\lfloor{Z'_i}\rfloor} K\left[f'_i, \textrm{nr}(f'_{i,n})\right]~,
\end{equation}
with $Z'_i = 2 \pi f'_i a$ and $f'_{i,n} = f'_i + n / P$.  The \Cstat is computed by convolving the $2\mathcal{F}(f)$ with $q(f)$ \citep{mes07}
\begin{equation}
	\mathcal{C}(f'_i) = 2\mathcal{F} \ast q(f'_i)~.
\end{equation}

We compute the higher criticism statistic as described in Section \ref{sec:generalHC} for the $2\mathcal{F}$ and the $\mathcal{C}$ values contained in a given search window and call the results $HC_{2\mathcal{F}}$ and $HC_\mathcal{C}$ respectively.  The noise distribution for $2\mathcal{F}$ is $\chi^2_4(0)$.  It is more complicated for $\mathcal{C}$, as the number of sidebands $M'_i = 2 \lfloor{Z'_i}\rfloor + 1$ depends on the frequency $f'_i$.  Hence, the p-values for each $\mathcal{C}(f'_i)$ are calculated with respect to the $\chi^2_{4M'_i}$ noise distribution (four degrees of freedom for each sideband summed).

To compare the performance of the three detection statistics $\mathcal{C}$, $HC_{2\mathcal{F}}$ and $HC_\mathcal{C}$, we compute thresholds with a false alarm rate $\alpha = 0.01$.  We have $n$ $2\mathcal{F}$ and $\mathcal{C}$ values, one for each frequency bin, and $N_w$ $HC_{2\mathcal{F}}$ and $HC_\mathcal{C}$ values, one for each search window.  Thresholds are found by insisting that noise-only data fall below the threshold in every bin or window with probability at least $1-\alpha$ [e.g., \citet{wet12}],
\begin{equation}
	\label{eq:Cthresh} \left\{P\left[\mathcal{C} < \mathcal{C}_{th}(\alpha)\right]\right\}^{n} \ge 1 - \alpha~,
\end{equation}
\begin{equation}
	\label{eq:HCthresh} \left\{P\left[HC_{2\mathcal{F}, \mathcal{C}} < g(n_\mathrm{w}, \alpha)\right]\right\}^{N_\mathrm{w}} \ge 1 - \alpha~,
\end{equation}
where $\mathcal{C}_{th}(\alpha)$ is the \Cstat threshold.  The $HC$ threshold $g(n_\mathrm{w}, \alpha)$ is determined from Monte-Carlo simulations, as described in Section \ref{sec:detectthresh}.  We integrate the \Cstat probability distribution to find $\mathcal{C}_{th}(\alpha)$ and use the result to compute the threshold noncentrality parameter $\lambda_{th}(\alpha, \delta)$ for false dismissal rate $\delta = 0.1$ from
\begin{equation}
	\label{eq:Cstatthresh}\int_{\mathcal{C}_{th}(\alpha)}^\infty dx \;\; F(x; 4M'_{i,max}, 0)  = \alpha / n~,
\end{equation}
\begin{equation}
	\label{eq:Cstatfalsealarm}\int_0^{\mathcal{C}_{th}(\alpha)} dx \;\; F\left[x; 4M'_{i,max}, \lambda_{th}(\alpha, \delta)\right]  = \delta~,
\end{equation}
where $F(x; k, \lambda)$ is the probability density function of the $\chi^2_k(\lambda)$ distribution and $4M'_{i,max}$ is the maximum number of sidebands for all $f'_i$ in the window.

To simulate the gravitational wave search, we choose $\rho^2_0$ and draw $2\mathcal{F}$ values from the $\chi^2_4[\rho^2(f'_i)]$ distribution for each frequency bin $f'_i$.  The search windows are constructed to ensure the entire sideband structure lies completely within a single window.  Sidebands are also present in adjacent windows, however we focus only on a single window, assuming (conservatively) that all other windows contain only noise.  Equation (\ref{eq:HCthresh}) makes the same assumption when computing the detection thresholds against which $HC_{2\mathcal{F}}$ and $HC_\mathcal{C}$ are compared.  We conduct a 100 Monte-Carlo simulations for each $\rho^2_0$ and find the fraction where the computed statistic lies above its threshold value, which we call the detection rate.

\begin{figure}
	\plotone{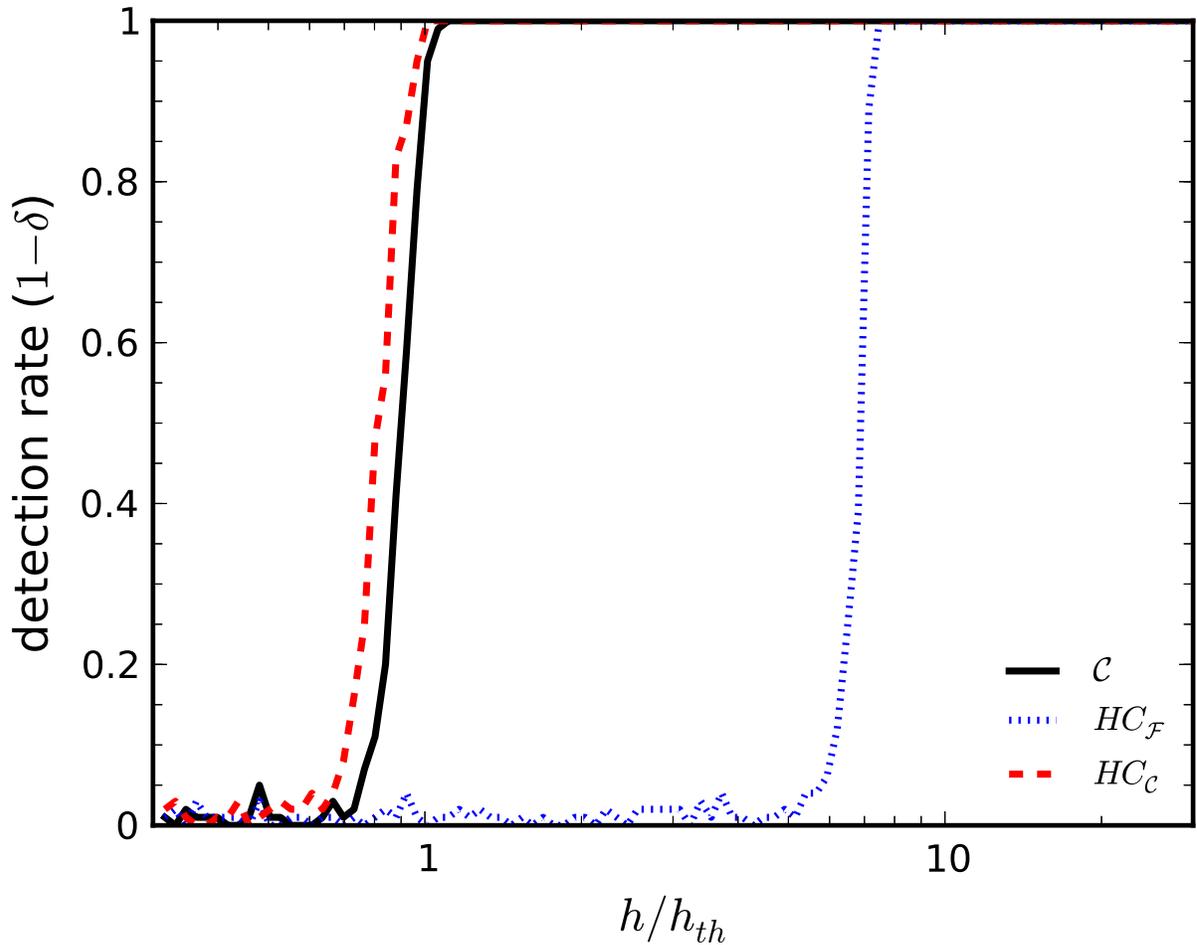}
	\caption{Monte-Carlo detection rate ($1-\delta$), equal to the fraction of simulations where the detection statistic lies above the detection threshold, as a function of wave strain, for the \Cstat (solid black), $HC_\mathcal{C}$ (dashed red), and $HC_{2\mathcal{F}}$ (dotted blue).  \label{fig:binary}}
\end{figure}

Figure \ref{fig:binary} displays detection rates for $\mathcal{C}$, $HC_{2\mathcal{F}}$, and  $HC_\mathcal{C}$ as functions of increasing wave strain $h / h_{th}$, where $h_{th}$ is defined as the wave strain corresponding to $\lambda_{th}$.  For $\delta = 0.1$ (detection rate of 90\%), $HC_\mathcal{C}$ detects wave strains 6\% smaller than $\mathcal{C}$ alone and $HC_{2\mathcal{F}}$ detects wave strains $\sim 7$ times greater than $\mathcal{C}$.  Additionally, higher criticism on $2\mathcal{F}$ compared to $2\mathcal{F}$ provides a advantage comparable to $HC_\mathcal{C}$ over $\mathcal{C}$ (not displayed in Figure \ref{fig:binary}).  The results in Figure \ref{fig:binary} assume perfect knowledge of the binary orbital period $P$.  In summary, therefore, second-pass higher criticism on $\mathcal{C}$ slightly boosts the sensitivity of the \Cstat, while second-pass higher criticism on $2\mathcal{F}$ is not competitive.  Both higher criticism statistics, in particular $HC_{2\mathcal{F}}$, are more robust than $\mathcal{C}$, as demonstrated in the next section.

\subsection{Robustness of higher criticism}\label{sec:robustness}
One significant advantage of higher criticism is its robustness when applied to signals that differ from the assumed form, e.g. due to phase wandering.  In this section, we explore the performance of the statistics $\mathcal{C}$, $HC_{\mathcal{C}}$, and $HC_{2\mathcal{F}}$ when the form of the signal remains unaltered but the true signal parameters differ from those assumed in the search.
\textbf{}
For a comb search targeting a binary source, the orbital period $P$ determines the sideband spacing of the \Fstat in the frequency domain.  An error $\Delta P$ between the assumed and true value of $P$ produces a comb that does not coincide with the actual sidebands.  The \Cstat sums noise at each sideband whose frequency does not match a true sideband location, lowering the sensitivity.  In contrast, $HC_{2\mathcal{F}}$ only assumes that all sidebands are located within a particular frequency window.  It does not rely on knowledge of the precise location of each sideband and consequently is unaffected by an error in $P$.  Both $HC_\mathcal{C}$ and the \Cstat lose sensitivity as $\Delta P$ increases, however one would expect the relative advantage of $HC_\mathcal{C}$ compared to $\mathcal{C}$ is expected to remain.

Observations of the orbital period of Sco X-1 have reported conflicting values for $P$.  \citet{got75} measured $P = 68023.8 \pm 0.09$ s from archival optical observations.  Recently, \citet{van03} measured $P = 68163.6 \pm 8.6$ s with the Rossi X-ray Timing Explorer, but did not observe any signification periodicity at or near the previously reported value of 68023.8 s.  Other LMXBs have larger uncertainty in their orbital parameters [e.g., see Tables 2--4 in \citet{wat08}].  In light of the potential uncertainty in $P$, the robustness of higher criticism is valuable.

\begin{figure}
	\plotone{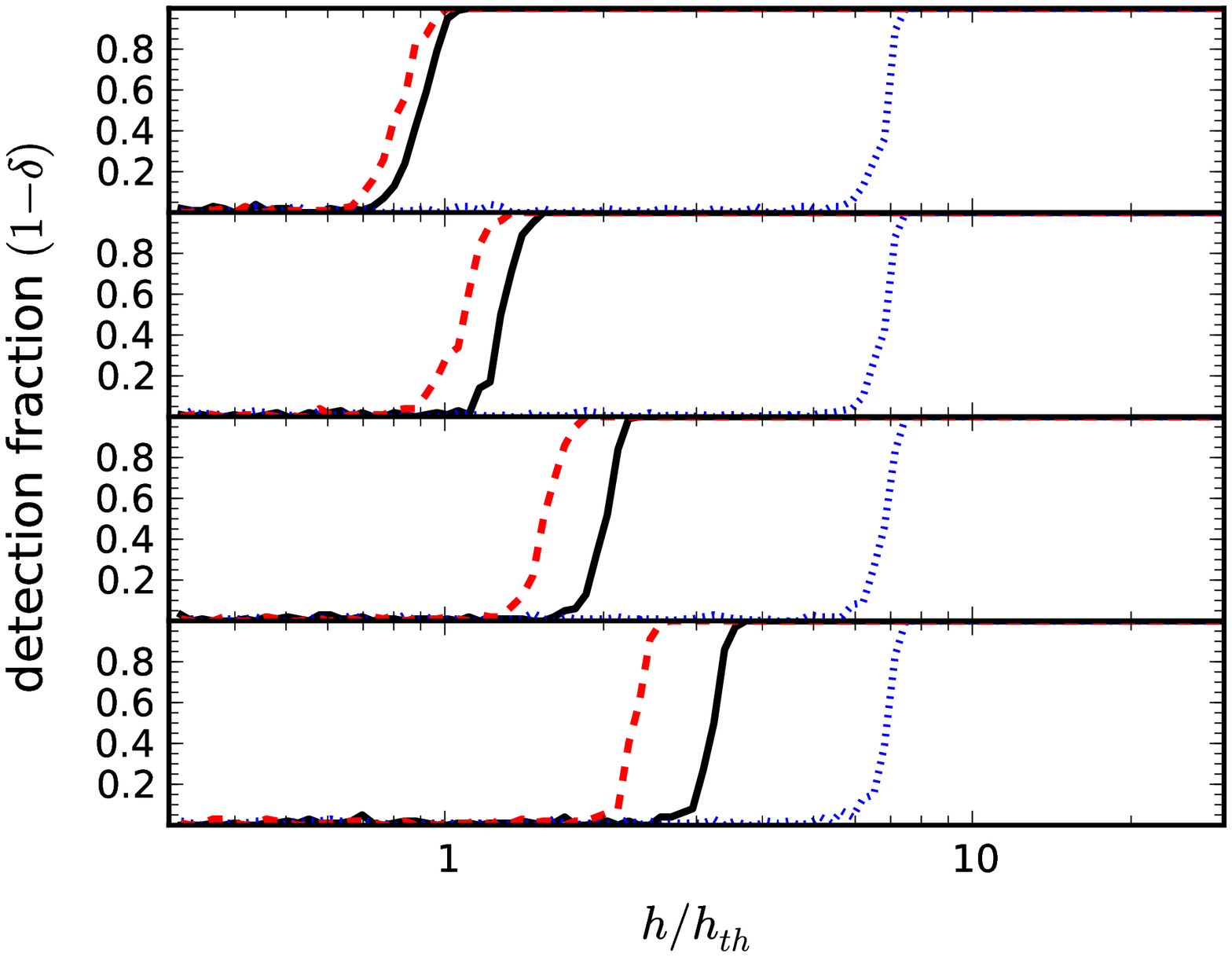}
	\caption{Monte-Carlo detection rate ($1-\delta$), equal to the fraction of simulations where the detection statistic lies above detection threshold, as a function of wave strain for the \Cstat (solid black), $HC_\mathcal{C}$ (dashed red), and $HC_{2\mathcal{F}}$ (dotted blue).  The error $\Delta P$ between assumed and true binary orbital periods increases from top to bottom: $\Delta P = 0, 1, 3, 9$ s.  \label{fig:binaryrobustness}}
\end{figure}

We repeat the simulations described in Section \ref{sec:binaryHC} with the addition of an error $\Delta P$ between the assumed value of $P$ used to construct the comb $q(f)$, and the true value used to generate the noncentrality parameters $\rho^2(f)$ for the \Fstat at every sideband.  Figure \ref{fig:binaryrobustness} displays detection rates with $\mathcal{C}$, $HC_{2\mathcal{F}}$, and  $HC_\mathcal{C}$ as $\Delta P$ increases.  As expected, $HC_{2\mathcal{F}}$ is unaffected by $\Delta P$ because higher criticism does not use any information about the sideband locations.  $\mathcal{C}$ and $HC_\mathcal{C}$ both lose sensitivity as $\Delta P$ increases, but the relative advantage of $HC_\mathcal{C}$ over $\mathcal{C}$ increases from 6\% when $\Delta P = 0$ to 15\% for $\Delta P = 1$ s, 21\% for $\Delta P = 3$ s, and 29\% for $\Delta P = 9$ s.  The difference between $\mathcal{C}$ and $HC_{2\mathcal{F}}$ decreases from a factor of 7 when $\Delta P = 0$ to 5 for $\Delta P = 1$ s, 3 for $\Delta P = 3$ s, and 2 for $\Delta P = 9$ s.  When $\Delta P = 0$, the \Cstat peaks at the signal frequency $f_0$.  For $\Delta P > 0$, the error between the assumed and actual location of the sidebands accumulates to the point where the outermost sidebands no longer coincide with the assumed comb.  This reduces the total number of signal sidebands summed by the \Cstat, reducing its maximum amplitude (at $f_0$).  Additionally, the \Cstat peak broadens compared with $\Delta P = 0$; this maximum amplitude requires less signal bins summed and can now be reached at a number of the central sidebands.  This situation moves closer to multiple, equal-strength signals, the problem for which higher criticism is designed, explaining the increasing advantage of $HC_\mathcal{C}$ over $\mathcal{C}$.  In contrast, the \Cstat requires a single source above threshold for detection.  Interestingly, the sensitivity decrease of $\mathcal{C}$ and $HC_\mathcal{C}$ stalls for 9 s $\lesssim \Delta P \lesssim$ 1000 s, when the accumulated error at the outermost sidebands is at least the sideband spacing.  At this point, sidebands begin to overlap with other parts of the comb, shifted by integer multiples of the sideband spacing.  As $\Delta P$ increases beyond 9 s, the number of sidebands correctly located by the template continues to decrease but is compensated for by a corresponding increase in the number of integer overlap sidebands.

\section{Phase wandering}\label{sec:wandering}
An accreting neutron star whose phase wanders in response to a variable accretion torque emits gravitational wave power in many frequency bins.  Resampling is difficult as the phase model is usually unknown, e.g. there may be an offset between the radio/X-ray ephemeris and gravitational wave signal.  Higher criticism can handle phase wandering robustly.

In its simplest form, a semi-coherent \Fstat combines $N$ coherent time intervals of equal length according to,
\begin{equation}
	\mathcal{F}_{sc}(f) = \sum_{i=1}^{N} \mathcal{F}_i(f)~,
\end{equation}
where $\mathcal{F}_i$ is the usual \Fstat for the $i$-th interval.  This method assumes a monochromatic source or, with some modification, a source whose phase evolves in a known way.  \citet{abb08a} described three semi-coherent methods to search for periodic gravitational waves in LIGO data.  The methods cannot be applied to a source whose phase wanders unpredictably, e.g., due to accretion torque.  We apply higher criticism to reanalyze detection statistics from multiple time intervals with this situation in mind.

We define $n_\mathrm{bins}$ to be the number of frequency bins and $n_\mathrm{GW}$ to be the number of bins containing a gravitational wave signal, assuming for simplicity that all signal bins have equal amplitude.  In practical terms, this situation corresponds to $n_\mathrm{GW}$ distinct sources of similar strength, or $n_\mathrm{GW}$ orbitally modulated sidebands of a source in a binary system.  We also assume that $n_\mathrm{bins}$ and $n_\mathrm{GW}$ are the same in each time interval, therefore $\epsilon = n_\mathrm{GW} / n_\mathrm{bins}$ is constant.  We allow for the signal to jump arbitrarily between bins (not necessarily adjacent ones) from one time interval to the next, e.g., due to phase wandering.

For each interval, we obtain a $2\mathcal{F}$ value for each of the $n_\mathrm{bins}$ frequency bins.  The $n_\mathrm{GW}$ signal bins follow a $\chi^2_4(\rho^2)$ distribution, the remainder obey a $\chi^2_4(0)$ distribution.  Combining all $2\mathcal{F}$ values for the $N$ intervals gives a total of $n = N n_\mathrm{bins}$ values.  Converting from $\epsilon$ and $n$ to $r$ and $\beta$ through equations (\ref{eq:parameps}) and (\ref{eq:paramchi2}), we use the theoretical detection boundary [equation (\ref{eq:detection_boundary})] to find $\rho^2_{HC}(N, n_\mathrm{GW}, n_\mathrm{bins})$, the minimum noncentrality detectable with $HC$ over $N$ time intervals, if $n_\mathrm{GW}$ of the $n_\mathrm{bins}$ bins contain signal.  This can be converted to a wave strain using equation (\ref{eq:rho2h0relation}).  The \Fstat threshold noncentrality $\rho^2_\mathcal{F}$ for a single interval is calculated similar to the \Cstat [equations (\ref{eq:Cstatthresh}) and (\ref{eq:Cstatfalsealarm})] \citep{wet12},
\begin{equation}
	\label{eq:Fstatthresh}\int_{\mathcal{F}_{th}}^\infty dx \;\; F(x; 4, 0)  = \alpha / n ~,
\end{equation}
\begin{equation}
	\label{eq:Fstatfalsealarm}\int_0^{\mathcal{F}_{th}} dx \;\; F\left(x; 4, \rho^2_{\mathcal{F}}\right)  = \delta^{1/n_\mathrm{GW}}~.
\end{equation}
The semi-coherent \Fstat threshold $\rho^2_{\mathcal{F}_{sc}}$ can also be calculated from equations (\ref{eq:Fstatthresh}) and (\ref{eq:Fstatfalsealarm}), substituting $4N$, rather than 4, degrees of freedom and assuming a monochromatic source.  The semi-coherent \Fstat cannot be applied to source whose phase wanders unpredictably and the threshold reverts to $\rho^2_\mathcal{F}$.

Comparing detection thresholds for $HC$ and $\mathcal{F}$ is difficult because the detection boundary for $HC$, derived in the limit $n\to\infty$, is not equivalent to the \Fstat thresholds, computed for a specific false alarm and false dismissal probability.  Furthermore, the $HC$ detection boundary underestimates the threshold for finite $n$ (see Section \ref{sec:detectthresh}).  We therefore recalibrate $\rho^2_{HC}$ by a constant such that $\rho^2_{HC}(N = 1, n_\mathrm{GW}=1, n_\mathrm{bins}) = \rho^2_\mathcal{F}$, thereby arranging that both $HC$ and $\mathcal{F}$ have equal power when $N = n_{GW} = 1$.  To test this recalibration, we conduct Monte-Carlo simulations to determine $\rho^2_{HC}$ for $\alpha = 0.01$, $\delta = 0.1$ when $N = n_\mathrm{GW} = 1$ for $n_\mathrm{bins}=10^3, 10^4, 10^5, 10^6$.  We find that the Monte-Carlo $\rho^2_{HC}$ agrees with $\rho^2_\mathcal{F}$ computed from equations (\ref{eq:Fstatthresh}) and (\ref{eq:Fstatfalsealarm}) to within 1\%.  However, it is important to mind the difference between each threshold when interpreting the following results.

\begin{figure}
	\plotone{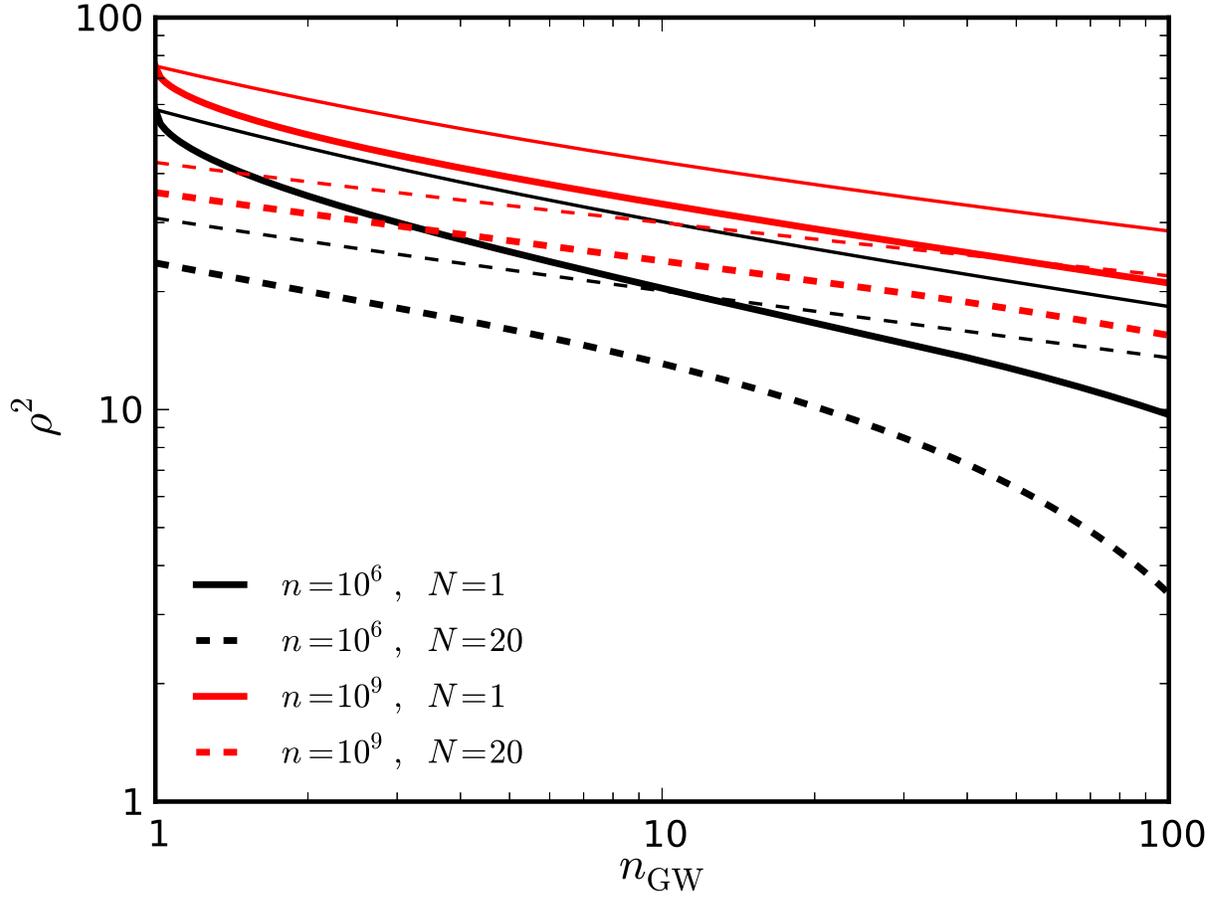}
	\caption{Noncentrality thresholds $\rho^2_{HC}$ (thick curves) and $\rho^2_\mathcal{F}$ (thin curves) as functions of $n_\mathrm{GW}$ for $(n_\mathrm{bins}, N)$ = $(10^6, 1)$, $(10^6, 20)$, $(10^9, 1)$, $(10^9, 20)$\label{fig:HCvsF_nGW}}
\end{figure}

\begin{figure}
	\plotone{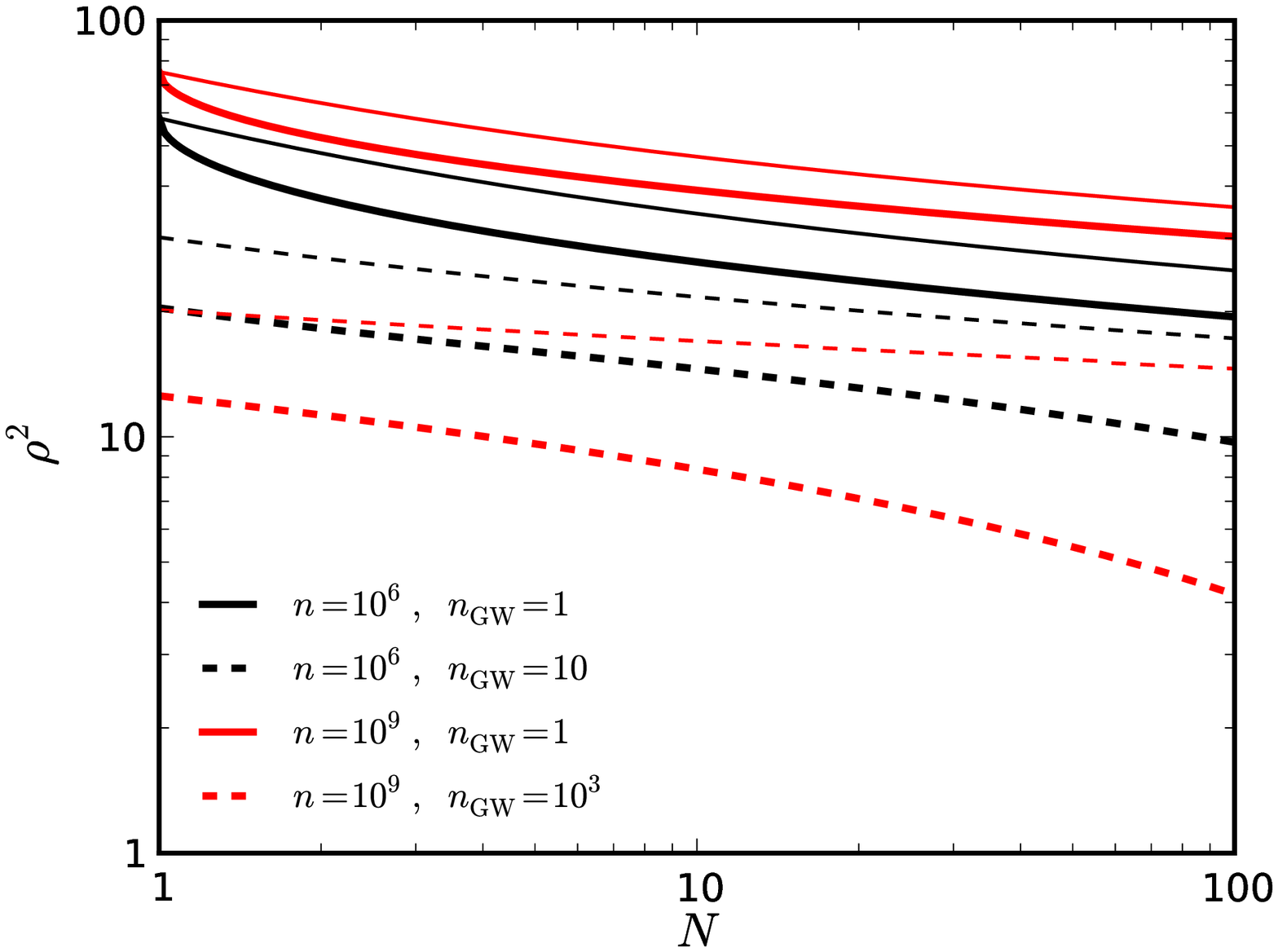}
	\caption{Noncentrality thresholds $\rho^2_{HC}$ (thick curves) and $\rho^2_\mathcal{F}$ (thin curves) as functions of $N$ for $(n_\mathrm{bins}, n_\mathrm{GW})$ = $(10^6, 1)$, $(10^6, 10)$, $(10^9, 1)$, $(10^9, 10^3)$\label{fig:HCvsF_N}}
\end{figure}

Figures \ref{fig:HCvsF_nGW} and \ref{fig:HCvsF_N} display $\rho^2$ required for detection with $HC$ (thick curves) and $\mathcal{F}$ (thin solid and dashed curves) for different combinations of $N$, $n_\mathrm{GW}$, $n_\mathrm{bins}$, and $\alpha = 0.01$, $\delta = 0.01$.  Figure \ref{fig:HCvsF_nGW} displays the thresholds $\rho^2_{HC}$ and $\rho^2_\mathcal{F}$ as functions of $n_\mathrm{GW}$ for $(n_\mathrm{bins}, N)$ = $(10^6, 1)$, $(10^6, 20)$, $(10^9, 1)$, $(10^9, 20)$.  Figure \ref{fig:HCvsF_N} displays $\rho^2$ as a function of $N$ for $(n_\mathrm{bins}, n_\mathrm{GW})$ = $(10^6, 1)$, $(10^6, 10)$, $(10^9, 1)$, $(10^9, 10^3)$.  For example, the search described in Section \ref{sec:binary} has $n_\mathrm{bins} \sim 10^9$ and $n_\mathrm{GW} \sim 10^4$ (the number of orbital sidebands).  The semi-coherent $\mathcal{F}_{sc}$ is, in general, the most sensitive but as described above we are most interested in sources with unknown phase wandering, for which semi-coherent searches are not suitable.  In all cases shown, $HC$ outperforms $\mathcal{F}$.  In Figure \ref{fig:HCvsF_nGW}, we observe reduction in $\rho^2_{HC}$ by a factor of 1.5 to 2.9 for $n_\mathrm{GW}=10$ and 2.3 to 7.0 for $n_\mathrm{GW}=100$, compared to $\rho^2_{HC}$ when $n_\mathrm{GW}=1$.  In Figure \ref{fig:HCvsF_N}, we observe reduction in $\rho^2_{HC}$ by a factor of 1.4 to 2.2 for $N=10$ and between 2.1 and 3.0 for $N=100$, compared to $N=1$.

The robustness of $HC$ described in Section \ref{sec:wandering} allows one to combine data from multiple time intervals to increased the sensitivity, even for a source whose phase wanders unpredictably.  In principle, one might improve detectability while remaining robust by including some simple information about the source frequency evolution, e.g. limiting wandering to some physically motivated range.  In Figures \ref{fig:HCvsF_nGW} and \ref{fig:HCvsF_N}, we assume no correlation, allowing the unlikely possibility that the frequency wanders across the entire range covered by the search.  One simple method is to divide the frequency range into intervals wide enough to contain the source throughout the observation and calculate $HC$ for each interval.

\section{Discussion}\label{sec:discussion}
Higher criticism is a recently formulated statistical method designed to detect the presence of a sparse collection of signals too weak to be detected individually.  In this paper, we apply higher criticism as a second pass to reanalyze gravitational wave search statistics and explore the feasibility of applications in two contexts: a targeted binary search (e.g. LMXB) and a phase-wandering source (e.g., glitching pulsar).

One advantage of higher criticism is its robust nature; it accommodates deviation from the expected phase evolution that hamper other search statistics.  However, there is a trade-off.  Higher criticism neglects some of the additional information used by other search methods which increase their performance under ideal conditions when the phase model is known.

In Section \ref{sec:binary}, we compare the performance of higher criticism and the \Cstat for a targeted binary search.  The \Cstat is more sensitive than higher criticism applied to \Fstat values by a factor $\sim 7$, however, higher criticism applied to \Cstat values gives a second-pass improvement of 6\% over the \Cstat.  Furthermore, higher criticism is more robust to an error $\Delta P$ between the true and assumed (from observation) binary orbital period.  While the absolute sensitivity of both $\mathcal{C}$ and $HC_\mathcal{C}$ decrease as $\Delta P$ increases, the performance of $HC_\mathcal{C}$ relative to $\mathcal{C}$ increases to a sensitivity improvement of $16\%$ for $\Delta P = 1$ s and $29\%$ for $\Delta P = 9$ s.

In Section \ref{sec:wandering}, we consider a phase wandering source.  The robustness of $HC$ allows one to combine data from multiple time intervals and boost the sensitivity compared to a single interval, even for a unpredictable source.  The noncentrality threshold decreases by a factor $\gtrsim 1.4$ over 10 intervals and a factor $\gtrsim 2$ over 100 intervals.

Another candidate for higher criticism is an all-sky search for unknown periodic sources.  (We tested this in Appendix \ref{sec:all_sky}).  The results indicate that higher criticism provides no advantage over the \Fstat, the statistics perform similarly. This stems from the distribution of source amplitudes: if neutron stars are uniformly distributed across the Galaxy, the closest, strongest source dominates the detectability.  There is no advantage using higher criticism, which is designed to detect a group of signals, when there is effectively only one signal present.

While there is reason to be optimistic about possible applications of higher criticism, we draw attention to a number of significant concerns.  Chief among these, higher criticism relies on knowing the background noise distribution.  In this paper we adopted the common, simplifying assumption of stationary, Gaussian detector noise.  In reality, detector noise is more complicated.  However, it is well studied and modified forms of higher criticism have been developed for correlated noise \citep{hal10} or when the noise distribution is imperfectly known \citep{del11}.

Additionally, higher criticism only detects the presence of a group of signals.  It cannot directly identify an individual source, which remains the ultimate goal of gravitational wave detection.  However, given the cheap computational cost of applying higher criticism as a second pass to reanalyze already computed search statistics, and the advantages related to its robustness, it has the potential to complement and enhance existing and future searches, especially as a guide to where to look harder in parameter space for just-too-weak sources.

\acknowledgements
We thank L. Sammut, V. Dergachev, and members of the Continuous Wave Search Group of the LIGO Scientific Collaboration for helpful discussions.
MFB was supported by an Australian Postgraduate Award.
AM, AD and PH were supported by grants and fellowships from the Australian Research Council.

\appendix
\section{All-sky search}\label{sec:all_sky}
One might ask whether higher criticism can be applied profitably to all-sky searches.  After all, from an estimated Galactic population of $10^9$ neutron stars \citep{arn89}, only $\sim 2000$ pulsars have been discovered as radio sources, of which $\sim 10$\% have frequencies in the range where current interferometers are most sensitive, $f_0 > 100$ Hz \citep{man05}\footnote{http://www.atnf.csiro.au/research/pulsar/psrcat}.  Theories of neutron star quadrupoles suggest that only a handful of the observed pulsars have any chance of being detectable with current sensitivity limits \citep{mas12}, but many of the radio-quiet objects are much closer to the Earth than their radio-loud brethren.    All-sky searches cover a range of frequencies at many sky locations, hoping to detect one of these unknown sources.  A number of all-sky searches have been carried out \citep{abb05, abb07b, abb08a, abb09a, abb09b, abb09d}, but no detection has been announced.  The searches typically use the \Fstat, or related statistics.  Hence they are candidates for reanalysis with higher criticism.

In contrast to the other applications in this paper, the sources targeted by an all-sky search have a power law (non-uniform) wave strain distribution.  We follow \citet{cut05} and assume neutron stars are spread uniformly throughout the Galaxy.  Then the distance distribution separates into two parts: a local uniform three-dimensional distribution up to the thickness of the Galactic disk, and a uniform two-dimensional distribution beyond.  Taking the disk to be 600 pc thick and 10 kpc in diameter, the 3D and 2D distributions correspond to $D < 300$ pc and 300 pc $< D <$ 5 kpc respectively, where $D$ is the distance from Earth.  We further assume that all sources have the same intrinsic amplitude, so that $h_0$ at Earth depends on distance alone.  Changing variables from distance $D$ to noncentrality parameter $\rho^2$, [see equation (\ref{eq:noncentrality_paramater}); also \citet{cut05} for details] we arrive at the distribution,
\begin{equation}\label{eq:rho2dist}
	\sigma(\rho^2) = \Bigg\{ \begin{array}{ll} n_{\mathrm{3D}} [\rho^2]^{-5/2}, & \textrm{(3D)}~, \\ n_{\mathrm{2D}} [\rho^2]^{-2}, & \textrm{(2D)}~,\end{array}
\end{equation}
where $n_\mathrm{2D}$ and $n_\mathrm{3D}$ are normalization constants which depend on $h_{0}$, $S_h(f)$ and $T_{obs}$ through equation (\ref{eq:noncentrality_paramater}).

To illustrate, we consider the same search parameters as a completed LIGO all-sky search for periodic sources.  \citet{abb07b} searched the 10 hr of LIGO data with the best sensitivity from the second science run and computed the \Fstat for $1.6 \times 10^8$ frequency bins at each of $3 \times 10^4$ sky locations, a total of $n = 5 \times 10^{12}$ \Fstat values.  We estimate the number of neutron stars with frequencies that fall in the range of the search (160-728.8 Hz) to be $10^7$ (of an estimated Galactic population of $10^9$).  Under these assumptions, each of the $5 \times 10^{12}$ frequency bins is equally likely to contain some signal, with $\epsilon = 2 \times 10^{-6}$.

We compute the Monte-Carlo higher criticism detection rate as a function of $\rho^2_{max} / \rho^2_{\mathcal{F}}$, where $\rho^2_{\mathcal{F}}$ is the noncentrality threshold for a \Fstat detection with 1\% false alarm rate and 10\% false dismissal rate [see equations (\ref{eq:Fstatthresh}) and (\ref{eq:Fstatfalsealarm})].  We can then compare the noncentrality of the brightest source, $\rho^2_{max}$, required for detection with higher criticism with the equivalent quantity for a \Fstat search.  To save computation, we scale simulations from $n=5 \times 10^{12}$ to $n' = 10^6$ by keeping parameters $r$ and $\beta$ constant in equations (\ref{eq:parameps}) and (\ref{eq:paramchi2}) and converting from $\epsilon$ and $\rho^2$ for $n=5\times 10^{12}$ to $\epsilon'$ and ${\rho^2}'$ for $n'=10^{6}$.

Signal values are drawn from a $\chi_4^2({\rho^2}')$ distribution with noncentrality parameter ${\rho^2}'$ generated according to equation (\ref{eq:rho2dist}).  To fix $n_\mathrm{2D}$ and $n_\mathrm{3D}$ we specify $\rho^2_{max}$ for a source at a distance of 10 pc.  The remaining $\mathcal{F}$ values, containing only noise, are drawn from the $\chi_4^2(0)$ distribution, and we compute $HC$ for the \Fstat values.  For each $\rho^2$, we run 100 simulations and the detection rate equals the fraction of simulations with $HC$ or $\mathcal{F}$ above their corresponding thresholds for detection.  Finally, we convert noncentrality $\rho^2$ to wave strain $h$ through equation (\ref{eq:rho2h0relation}) to plot a more meaningful quantity.

\begin{figure}
	\plotone{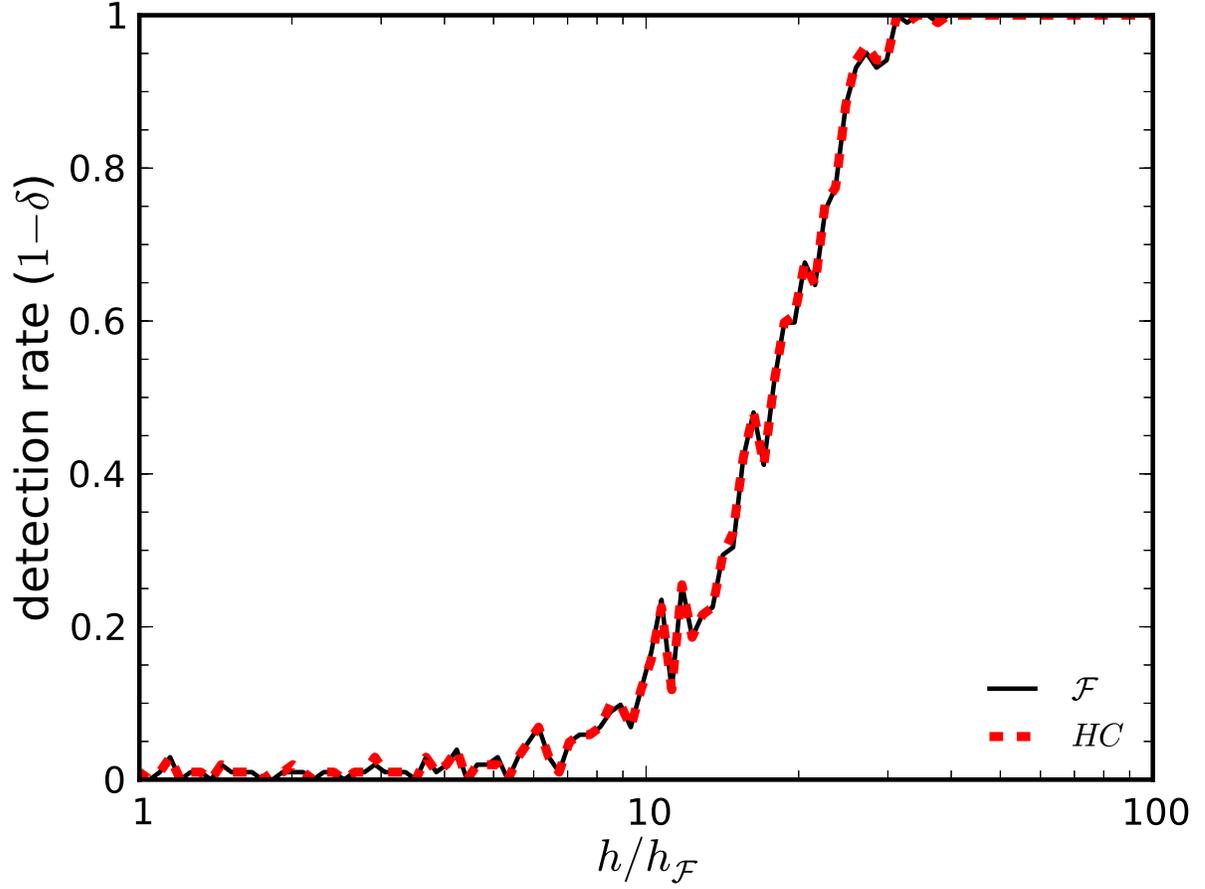}
	\caption{Monte-Carlo detection rate with higher criticism (thick dashed red) versus \Fstat (thin solid black) as a function of $h / h_\mathcal{F}$, where $h$ is the wave strain of a source located at 10 pc and $h_\mathcal{F}$ is the strain required by a source at 10 pc for detection with the \Fstat.\label{fig:allsky}}
\end{figure}

Figure \ref{fig:allsky} displays the Monte-Carlo detection rate as a function of $h / h_{\mathcal{F}}$, where $h$ is the wave strain for a source at 10 pc, corresponding to $\rho^2_{max}$, normalized by $h_{\mathcal{F}}$, the wave strain required for detection with the \Fstat at 10 pc.  Figure \ref{fig:allsky} shows that higher criticism provides no advantage over the \Fstat; they perform almost identically.  The fluctuations in Figure \ref{fig:allsky} relate to the randomly sampled source distribution.  Despite the large number of sources, $\sigma(\rho^2)$ in equation (\ref{eq:rho2dist}) falls off too fast with $\rho^2$ and effectively only the brightest source, rather than the collection of sources, is detected.  To test this we repeated the above simulations using only the single strongest source (replacing the rest with noise) and found the same result.  The similar performance of $\mathcal{F}$ and $HC_\mathcal{F}$ agrees with Section \ref{sec:wandering}, where Monte-Carlo simulations found $HC$ and $\mathcal{F}$ had equal thresholds when only one source was present.  Therefore, for an all sky search, higher criticism does not provide any benefit over a first pass \Fstat search.

\end{document}